\documentclass[epsf,a4wide]{article}
\usepackage{verbatim}
\usepackage{amsmath}
\usepackage{amssymb}

\large\normalsize \advance\textwidth 4cm \advance\textheight 2cm \advance\oddsidemargin -3cm
\advance\evensidemargin -1cm \advance\topmargin -3cm
\bigskip
\bigskip

\newcount\timehh\newcount\timemm
\timehh=\time \divide\timehh by 60 \timemm=\time \count255=\timehh\multiply\count255 by -60 \advance\timemm by
\bigskip
\bigskip
\bigskip
\bigskip
\begin{document}
 \centerline{\large\textbf {The Hyperbolic Bloch Equations of General Relativity}}
\bigskip
\centerline{Andrew Farley}
\centerline{ansjf2@gmail.com; ORCHID: 0000-0002-9991-7745}
\centerline{Guildford, Surrey, United Kingdom}
\bigskip
\bigskip
\centerline{\textbf{Abstract}}
\bigskip

New equations are derived which describe the evolution in curved spacetime of null geodesics with non-zero (complex) shear $\sigma$ and twist $\omega$ rates resembling Grishchuk's squeezed states evolution equations from inflationary cosmology. A ``squeeze" angle $\phi$ (obtained from the direction of the major axis of the elliptical cross section of the congruence and the direction of the shear rate), an ellipse axis ratio parameter $w$ and a rotation angle $v$ are the primary variables. Interpreting $\phi$ as a polar angle and $w$ as a radial distance, we obtain a mapping to points on the upper sheet, $H_{2}^{+}\,,$ of a two-sheet hyperboloid, establishing the connection between gravitational optics and hyperbolic geometry. Points on $H_{2}^{+}$ trace out paths evolving according to hyperbolic Bloch equations, similar to the optical Bloch equations, which can also be represented as a Schr\"{o}dinger-like equation with a non-Hermitian Hamiltonian. A single vector equation on $H_{2}^{+}$ describes the precession of hyperbolic Bloch vectors about a rotation or birefringence vector on $H_{2}^{+}\,,$ analogous to the precession of Bloch vectors on the Bloch sphere or Stokes vectors on the Poincar\'{e} sphere. Tidal gravitational effects and a non-zero twist $\omega$ contribute to the precession of hyperbolic Bloch vectors.

\section{Introduction}

Geometrical optics, as a leading-order approximation to the solution of Maxwell's equations in curved spacetime, underpins many of the theoretical approaches to gravitational lensing. Light rays are described as the infinitely high-frequency limit of electromagnetic waves whose null propagation vectors $k^{\mu}\,,$ ($k^{\mu}g_{\mu\nu}k^{\nu}\,=\,0\,,$ where $g_{\mu\nu}$ is the spacetime metric) form a congruence of, and are tangent to, affinely parameterised null geodesics. Wavelengths are much shorter than characteristic length scales, like spacetime and wavefront curvature scales, which makes geometrical optics a good approximation for observational purposes. Maxwell's equations in this approximation also imply that polarisation vectors are parallelly transported along the rays and the photon number is conserved along the rays such that the square field amplitude is inversely proportional to the transverse area of the beam. Wavefronts are null hypersurfaces of constant phase $S$ orthogonal to null geodesic tangent vectors $k^{\mu}$ which are generators of the wavefronts. The eikonal equation for $S$ follows from $k^{\mu}$ being a null vector and the geodesic equation $k^{\mu}\nabla_{\mu}k^{\nu}\,=\,0$ follows since $k^{\mu}$ is a gradient: $k_{\mu}\,=\,\partial_{\mu}S\,$. On caustics of the wavefront, however, the geometrical optics approximation breaks down and the magnification factor is infinite for point sources. Higher-order geometrical optics or wave optics effects become important when wavelengths are comparable to or larger than typical spacetime and/or wavefront curvature length scales meaning diffraction and interference effects need to be taken into account. Astrophysical point sources of electromagnetic waves are, of course, only an idealisation. Taking more physical higher-order effects into account in gravitational lensing means the magnification along the rays is now finite.

Below we introduce alternative gravitational optics equations which describe the evolution of light rays, modelled as null geodesics, in a general curved spacetime. Rather than the geometry of polarisation states in curved spacetime, which we consider elsewhere, our focus is on the geometry of images from astrophysical sources. An axis ratio $w$ with respect to elliptical (due to tidal gravitational effects) cross sections of the congruence of rays from an astrophysical source and a ``squeeze" angle $\phi$ are introduced, both of which vary along the rays of the congruence with its affine parameter $\lambda\,$. Just as we expect the shape and orientation of the electric field's polarisation ellipse to change continuously as light traverses through a stressed medium, the shape and orientation of the images of astrophysical objects changes due to local and non-local matter. To complete the squeezed states analogy, a rotation angle $v$ is also introduced but not explored further. The $w\,,$ $\phi$ and $v$ parameters have evolution equations driven by the (complex) shear $\sigma$ and twist rate $\omega$ optical scalars. As is well known, the shear rate is itself driven by the tidal gravitational effects of distant matter through the $\Psi_{0}$ Weyl scalar. We suggest that the presence of shear may be indicative of the higher-order geometrical optics expansion of the electric field in curved spacetime. 

Illustrating a certain departure from leading-order geometrical optics described by null hypersurfaces of constant phase, where $k^{\mu}$ is a gradient, we assume that the null geodesics are twisting. Near physically realistic rotating sources such as a Kerr black hole, light rays are expected to be twisted. Having a non-zero twist rate $\omega$ means that light rays do not collapse to points or lines of infinite magnification as with standard geometrical optics.

As we will demonstrate, our new gravitational optics equations establish connections between gravitational optics and the Bloch equations describing two-level atoms interacting with light or, alternatively, the equations describing the evolution of polarisation states of light in an anisotropic medium. In the two-level atom model, Bloch's spin vector approach, which was originally conceived in the context of nuclear magnetic resonance, is applied to optical resonance problems such that components of a fictitious electric pseudospin vector evolve in accordance with the optical Bloch equations. Mathematically, therefore, the two-level atom is equivalent to a spin-$\frac{1}{2}$ particle in an external magnetic field. Geometrically, Bloch vectors move along the surface of the Bloch sphere which represents states of a two-level system. Fully excited and ground states are repesented by the poles of the Bloch sphere. The electric pseudospin vector precesses about a rotation vector whose magnitude is the generalised Rabi frequency. In polarisation optics, Stokes parameters describe various polarisation states of light propagating in a medium, which have a representation on the Poincar\'{e} sphere. In a weakly inhomogeneous and anisotropic medium, Stokes polarisation vectors with end points on the Poincar\'{e} sphere precess about a birefringence vector for the medium. Similarly, we shall show that in general relativity, a single vector equation describes the motion of points on the upper sheet of a two-sheet hyperboloid $H_{2}^{+}$ which geometrically represents the states of light rays interacting with a gravitational field. As shown below, an analogous precession occurs about a hyperbolic Bloch vector with endpoints on $H_{2}^{+}$ driven by the Weyl scalar $\Psi_{0}$ and angular momentum in the light rays.

It is well known that the standard two-level atom interacting with an external, time-dependent electromagnetic field has an underlying $SU(2)$ Lie group structure. In our (bosonic) gravitational optics model the underlying group structure is $SU(1,1)\,$. This is closely related to the non-Hermitian $(H_{ij}\,\neq\,H_{ji}^{*})\,$, time dependent Hamiltonian we obtain in a Schr\"{o}dinger-like representation of the hyperbolic Bloch equations. Non-Hermitian Hamiltonians are now an important part of theoretical and experimental physics since the discovery that the eigenvalues of a non-Hermitian but space–time reflection invariant ($PT-$symmetric) Hamiltonian can be entirely real \cite{Bender}. A more physically motivated $PT$ symmetry, therefore, replaced the mathematical axiom in quantum mechanics that the Hamiltonian is Hermitian with real eigenvalues driving a unitary (probability conserving) time evolution. The non-Hermitian Hamiltonian we obtain below has components related to the magnitude of the shear rate of the congruence and a detuning function. It also has the form that meets all the conditions of $PT$ quantum mechanics \cite{JonesSmith}. As a $PT$-symmetric Hamiltonian, either all its eigenvalues are real (unbroken $PT$-symmetry) or some are real and some are complex (broken $PT$-symmetry). Varying the parameters of the Hamiltonian allows one to pass between unbroken and broken phases.

This paper is structured as follows. We firstly present the (Sachs) optical scalars \cite{Sachs} associated with a twisting null geodesic congruence of rays in the Newman-Penrose formalism \cite{NP} and their associated evolution equations. Cross sections of the congruence are elliptical when the (complex) shear rate $\sigma$ is non-zero. Describing the evolution of elliptical cross sections sets the scene for the subsequent derivation of our new gravitational optics evolution equations. We show that such equations for $w$ and $\phi$ can be combined into a single complex evolution equation. Connections with hyperbolic geometry and the underlying $su(1,1)$ Lie algebra are then established leading to the hyperbolic Bloch equations. A vector representation of the evolution equations enables us to derive the precession of a hyperbolic Bloch vector around a rotation or bifrefringence vector. Future directions for this work are then discussed.

We employ units where $G\,=\,c\,=\,1$ and denote the spacetime metric by $g_{\mu\nu}(x)$ with Greek indices denoting 4-dimensional metric components. A general spacetime is considered and no symmetries are assumed unless otherwise stated. Conventions for the optical scalars follow from the Newman-Penrose formalism \cite{NP}.

\section{Optical Scalar Equations}

We consider a thin congruence or pencil of null geodesics, $\mathcal{N}\,,$ whose tangent is a (past-directed) null vector $k^{\mu}\,=\,\frac{dx^{\mu}}{d\lambda}\,,$ where $\lambda$ denotes the (unobservable) affine parameter along the past light cone and $x^{\mu}(\lambda)$ denotes the path of the light rays. The expansion rate of $\mathcal{N}\,$, $\theta\,,$ characterised by $k^{\mu}$ is just the fractional rate of change of the infinitesimal cross-sectional area $\Delta A$ of $\mathcal{N}\,$ orthogonal to $k^{\mu}$ such that $\theta\;=\;\frac{\dot{\Delta A}}{\Delta A}\,=\,\nabla_{\mu}k^{\mu}\,,$ where an overdot denotes $k^{\mu}\nabla_{\mu}\,.$ The complex function $\sigma\,=\,-m^{\mu}m^{\nu}\nabla_{\nu}k_{\mu}\,=\,\mid\sigma\mid e^{i\varphi}$ denotes the shear rate of the congruence describing how tidal effects distort circular cross sections of $\mathcal{N}$ into ellipses, where $\varphi(\lambda)$ denotes the direction of the shear rate. However, $\sigma$ depends on the choice of scaling along $\mathcal{N}$ and the choice of null complex vectors $m^{\mu}$ ($\bar{m}^{\mu}$ its conjugate such that $m_{\mu}\bar{m}^{\mu}\,=\,1$) orthogonal to $k^{\mu}$ $(k^{\mu}m_{\mu}\,=\,0)$ constructed from a pair of spacelike unit vectors $e^{\mu}_{\pm}\,,$ such that $m^{\mu}\,=\,\frac{1}{\sqrt{2}}(e_{+}^{\mu}\,+\,ie_{-}^{\mu})\,,$ $e_{\pm}^{\mu} e_{\pm}^{\nu}g_{\mu\nu}\,=\,1\,,$ $e_{-}^{\mu} e_{+}^{\nu}g_{\mu\nu}\,=\,0\,.$ For an affinely parametrised congruence, then one can show that $k^{\nu}\nabla_{\nu}k^{\mu}\,=\,-\kappa\bar{m}^{\mu}\,-\,\bar{\kappa}m^{\mu}\,,$ where $\kappa\,=\,-m_{\nu}k^{\mu}\nabla_{\mu}k^{\nu}$ is a Newman-Penrose spin coefficient. Since $k^{\mu}m_{\mu}\,=\,0\,,$ then adopting two-dimensional ``screen" vectors $\{m^{\mu}\,,\,\bar{m}^{\mu}\}$ which are parallely transported null vectors along the rays $(k^{\mu}\nabla_{\mu}m^{\nu}\,=\,0)$ then $\kappa\,=\,0$ means and the rays follow geodesics $k^{\nu}\nabla_{\nu}k^{\mu}\,=\,0\,.$ 

The complex Newman-Penrose scalar $\rho\,=\,-m^{\mu}\bar{m}^{\nu}\nabla_{\nu}k_{\mu}$ can be written in terms of the expansion and twist rates of the congruence as $\rho\,=\,-\frac{1}{2}(\theta\,+\,i\omega)\,,$ where $\omega$ denotes the twist rate such that $\omega^{2}\,=\,2\nabla_{[\mu}k_{\nu]}\nabla^{\mu}k^{\nu}\,.$ In geometrical optics, $\mathcal{N}$ is hypersurface orthogonal ($\nabla_{[\mu}k_{\nu]}\,=\,0\,;$ i.e., $\omega\,=\,0$), which means $k_{\mu}$ can be expressed as the gradient of a scalar field which is constant on the wavefronts. We do not make that assumption here ($\omega\,\neq\,0$) so $\mathcal{N}$ is twisting or rotating. The twist can also be written
\begin{equation}\label{tw1}
\omega\;=\;\epsilon^{\mu\nu\sigma\rho}n_{\sigma}k_{\rho}\nabla_{\mu}k_{\rho}\quad,
\end{equation}
where $\epsilon^{\mu\nu\sigma\rho} $ denotes the totally anti-symmetric Levi-Civita tensor and $n^{\rho}$ is a null vector which completes the null tetrad $(k^{\mu}\,,\,n^{\mu}\,,\,m^{\mu}\,,\,\bar{m}^{\mu})$ with $n^{\mu}k_{\mu}\,=\,-1$ with all other contractions zero. The function $\rho$ satisfies the NP equation \cite{NP}
\begin{equation}\label{NP}
\dot{\rho}\;=\;\rho^{2}\,+\,\mid\sigma\mid^{2}\,+\,\Phi_{00}\quad,
\end{equation}
where $\Phi_{00}\,=\,\frac{1}{2}R_{\mu\nu}k^{\mu}k^{\nu}\,=\,4\pi T_{\mu\nu}k^{\mu}k^{\nu}$ is real with $R_{\mu\nu}$ the Ricci tensor and $T_{\mu\nu}$ the energy-momentum tensor of local matter. By the null energy condition, $\Phi_{00}\,\geq\,0\,$. Splitting eq.(\ref{NP}) into real and imaginary parts, then $\theta$ satisfies the Raychaudhuri equation and $\omega$ satisfies $\dot{\omega}\,=\,-\theta\omega\,.$ The solution for $\omega$ is $\omega\,=\,\frac{2L}{D_{a}^{2}}\,,$ where $L$ is a constant $(\dot{L}\,=\,0)$ related to the angular momentum in the beam, such that $\ddot{D}_{a}\,+\,(\mid\sigma\mid^{2}\,+\,\Phi_{00})D_{a}\,=\,\frac{L^{2}}{D_{a}^{3}}\,$. Thus, we take $L\,\neq\,0$ which means that cross sections of $\mathcal{N}$ do not collapse to a line or point as they do in geometrical optics \cite{perlick}. Intuitively, this is expected from the Raychaudhuri equation since $\omega^{2}\,>\,0$ can counteract the focusing effects of shear and matter.

The shear rate evolves along the null congruence according to $\dot{\sigma}\,+\,\theta\sigma\,=\,\Psi_{0}$ \cite{NP} where
\begin{equation}\label{OS4}
\Psi_{0}\;=\; C_{\mu\sigma\rho\nu}k^{\sigma}k^{\rho}\bar{m}^{\mu}\bar{m}^{\nu}\;=\;\Psi e^{i\mu}
\end{equation}
denotes one of the Weyl scalars constructed from the Weyl tensor, $C_{\mu\sigma\rho\nu}\,,$ which encompasses the tidal gravitational effects of distant matter. Naturally, $\Psi (\lambda)$ is the real magnitude of $\Psi_{0}$ whilst $\mu$ is a real phase describing the angular position of the deflector or lens relative to the light beam. The amplitude $\Psi$ is the ultimate source of anisotropic focusing, such that circular beam cross sections evolve into an elliptical shape \cite{Penrose}. From the Newman-Penrose equation for $\sigma$ and eq.(\ref{OS4}), the phase $\varphi$ evolves according to
\begin{equation}\label{OS5}
\Omega\;\equiv\;\dot{\varphi}\;=\;\frac{\Psi}{\mid\sigma\mid}\sin (\mu\,-\,\varphi)\quad.
\end{equation}
The magnitude of the shear evolves according to
\begin{equation}\label{OS5a}
\dot{\mid\sigma\mid}\,+\,\theta \mid\sigma\mid\;=\;\Psi\cos(\mu\,-\,\varphi)\quad,
\end{equation}
with solution
\begin{equation}\label{shear1}
\mid\sigma\mid\;=\;\frac{\sigma_{0}}{D_{a}^{2}(\lambda)}\,+\,\frac{1}{D_{a}^{2}(\lambda)}\int_{\lambda_{0}}^{\lambda}d\lambda'\,D_{a}^{2}(\lambda')\Psi (\lambda')\cos(\mu\,-\,\varphi)\quad,
\end{equation}
where $\sigma_{0}$ is a non-zero constant (i.e., independent of $\lambda$) and $\lambda_{0}$ is some fixed affine parameter.

\section{Elliptical Cross Sections}

We define a connection vector $\eta^{\mu}$ between neigbouring rays in the congruence which can be written in terms of the null tetrad basis as
$\eta^{\mu}\,=\,gk^{\mu}\,+\,hn^{\mu}\,+\,\bar{\eta}^{\mu}\,,$
where $g$ and $h$ are real functions and $\bar{\eta}^{\mu}$ denotes the connection vector projected onto a spacelike, two-dimensional screen orthogonal to the line-of-sight $\{e^{\mu}_{A}\}$ $(A\,=\,1,2)$
\begin{equation}\label{bareta}
\bar{\eta}^{\mu}\;=\;e^{\mu}_{A}\eta^{A}\;=\;h^{\mu}{}_{\nu}\eta^{\nu}\;=\;\zeta\bar{m}^{\mu}\,+\,\bar{\zeta}m^{\mu}\quad,
\end{equation}
where $\zeta$ is complex and $h^{\mu\nu}\,=\,h^{AB}e^{\mu}_{A}e^{\nu}_{B}\,=\,e_{+}^{\mu}e_{+}^{\nu}\,+\ e_{-}^{\mu}e_{-}^{\nu} $ is the screen projection operator. Hence, $\bar{\eta}^{\mu}$ represents the relative position on the screen of two light rays separated by $\eta^{\nu}\,.$ Evidently, $\bar{\eta}^{\mu}k_{\mu}\,=\,0$ since $e^{\mu}_{A}k_{\mu}\,=\,0\,.$ The components of $\eta^{A}$ are just the $(x\,,\,y)$ coordinates on the two-dimensional screen. Of primary interest is the evolution of the screen and the light rays in $\mathcal{N}$ intersecting it. Therefore, the functions $g$ and $h$ will play no further part in our analysis. A straightforward calculation gives the linear geodesic deviation equation for a null geodesic congruence \cite{Poisson}
\begin{equation}\label{screen}
\ddot{\eta}^{A}\;=\;\mathcal{R}^{A}{}_{B}\eta^{B}\quad,
\end{equation}
where
\begin{equation}\label{TM1}
  \mathcal{R}^{A}{}_{B}\;=\;\begin{pmatrix}
\Psi\cos\mu\,-\,\Phi_{00} & -\Psi\sin\mu \\
-\Psi\sin\mu & -\Psi\cos\mu\,-\,\Phi_{00}
\end{pmatrix}\quad,
\end{equation}
is known as the optical tidal matrix \cite{Seitz}. This matrix is evidently symmetric, ultimately due to the symmetries of the Riemann curvature tensor. Note that the affine parameter time derivative in eq.(\ref{screen}) is an ordinary derivative since $\eta^{A}$ behaves like a scalar under $\lambda$ derivatives.

Let $C_{\pm}\,>\,0$ denote the semi-major / minor axes of the elliptical beam cross section with area $\Delta A\,=\,\pi C_{+}C_{-}$ and define $w (\lambda)$ via the axis ratio
\begin{equation}\label{zeta4b}
  \frac{C_{+}}{C_{-}}\;\equiv\;e^{-w}\quad.
\end{equation}
Hence, $w\,\geq\,0$ is a measure of the ellipticity of the beam's cross section which vanishes for a circular cross section ($C_{+}\,=\,C_{-}$). Large distortions, therefore, imply $w\,\gg\,1\,.$ By construction, we orientate our elliptical image such that $C_{-}\,\geq\,C_{+}\,\geq\,0\,$. Making the $w\,\to\,-w$ transformation does not change the results below.

Points on an ellipse projected onto the screen and measured with respect to the reference ray or origin are defined by the Jacobi field $\eta^{A}$ satisfying eq.(\ref{screen}) 
\begin{equation}
\eta^{A}(\lambda)\;=\;C_{+}(\lambda)\cos t_{-}(\lambda)\,C_{+}^{A}\,+\,C_{-}(\lambda)\sin t_{-}(\lambda)\,C_{-}^{A}\quad,
\end{equation}
where $C_{+}^{A}\,=\,\cos t_{+} e_{+}^{A}\,+\,\sin t_{+} e_{-}^{A}$ and $C_{-}^{A}\,=\,-\sin t_{+} e_{+}^{A}\,+\,\cos t_{+} e_{-}^{A}$ ($m^{A}\,=\, 2^{-\frac{1}{2}}(e_{+}^{A}\,+ie_{-}^{A}) $) are unit orthonormal vectors in the directions of the semi-major and minor axes of the ellipse, respectively. The angle $t_{+}(\lambda)$ describes the orientation of the semi-major axis with respect to the $e_{+}^{A}$ direction, which changes as we move along the light rays. Whilst the $e_{\pm}^{A}$ basis vectors are parallely transported along the rays, the $C_{\pm}^{A}$ basis vectors are not since $t_{+}$ changes along the beam, as shown below. In fact, the basis $C_{\pm}^{A}$ is related to the $e_{\pm}^{A}$ basis by a rotation such that $M^{A}\,\equiv\,2^{-\frac{1}{2}}(C_{+}^{A}\,+iC_{-}^{A})\,=\,m^{A}e^{-it_{+}}\,.$

Individual rays in $\mathcal{N}$ intersect the two-dimensional screen at points $(x\,,\,y)$ which can be represented by a complex number $\zeta\,=\,\eta^{A}m_{A}\,=\,x\,+iy$ which evolves according to \cite{Penrose} (cf. eq.(\ref{bareta}))
\begin{equation}\label{zeta}
\dot{\zeta}\;=\;-\rho\zeta\,-\,\sigma\zeta^{*}\quad.
\end{equation}
Points on the screen can also be expressed as \cite{Penrose}
\begin{equation}\label{zeta3}
\zeta\;=\;e^{it_{0}}\left(\zeta_{1}e^{it}\,+\,\zeta_{2}e^{-it}\right)\quad,
\end{equation}
where $t_{0}$ is a constant phase, $t$ is a dummy variable ranging from $0$ to $2\pi$ such that
\begin{equation}
C_{\pm}\;=\;\sqrt{2}(\mid\zeta_{1}\mid\,\pm\,\mid\zeta_{2}\mid)
\end{equation}
and, with $\varphi_{1,2}$ as the real $\lambda$-dependent phases of the complex functions $\zeta_{1,2}(\lambda)\,,$
\begin{equation}\label{tpm}
t_{+}\;=\;t_{0}\,+\,\frac{1}{2}(\varphi_{1}\,+\,\varphi_{2})\quad,\quad t_{-}\;=\;t\,+\,\frac{1}{2}(\varphi_{1}\,-\,\varphi_{2})\quad.
\end{equation}
Eq.(\ref{zeta}) implies
\begin{equation} \label{zeta3b}
  \dot{\zeta}_{1}\;=\;-\rho\zeta_{1}\,-\,\sigma  e^{-2it_{0}}\zeta^{*}_{2}\quad,\quad\dot{\zeta}_{2}\;=\;-\rho\zeta_{2}\,-\,\sigma e^{-2it_{0}}\zeta^{*}_{1}\quad.
\end{equation}
Thus, with $\zeta_{1,2}\,=\,\frac{1}{2\sqrt{2}}(C_{+}\,\pm\,C_{-})e^{i\varphi_{1,2}}\,,$ eqs.(\ref{zeta3b}) can be written in the form \cite{perlick}
\begin{equation}\label{cpm1}
\dot{C}_{\pm}\,+\,i\dot{t}_{+}C_{\pm}\,+\,i\dot{t}_{-}C_{\mp}\;=\;\left(-\rho\,\mp\,\sigma e^{-2it_{+}}\right)C_{\pm}\quad.
\end{equation}
Its imaginary part gives
\begin{eqnarray}
\dot{t}_{+} & = & \frac{1}{2}\omega\,+\,\frac{\mid\sigma\mid(C_{+}^{2}\,+\,C_{-}^{2})\sin\phi}{C_{+}^{2}\,-\,C_{-}^{2}}\quad,\label{sq3}\\
\dot{t}_{-} & = & \frac{2\mid\sigma\mid C_{+}C_{-}\sin\phi}{C_{-}^{2}\,-\,C_{+}^{2}}\quad\label{sq3a}\quad,
\end{eqnarray}
where we defined $\phi\,=\,2t_{+}\,-\,\varphi\,.$ The real part gives
\begin{equation}\label{es1}
\theta\;=\;\frac{\dot{C}_{+}}{C_{+}}\,+\, \frac{\dot{C}_{-}}{C_{-}}\quad,\quad 2\mid\sigma\mid\cos\phi\;=\; \frac{\dot{C}_{-}}{C_{-}}\,-\, \frac{\dot{C}_{+}}{C_{+}}\quad.
\end{equation}
Evidently, if $\phi\,=\,0\,,$ $n\pi$ ($n$ an integer), then the phase $t_{-}$ does not change as we move along the ray but $t_{+}$ does due to the twist $\omega\,.$ Indeed, inserting eq.(\ref{tw1}) into eq.(\ref{sq3}), then in the absence of shear or $\phi\,=\,0\,,$ $n\pi\,,$ $t_{+}$ evolves in a manner similar to the rotation of the polarisation vector known as gravitational Faraday rotation \cite{Yoo}\cite{shoom}.

As an aside, polar coordinates $(r\,,\,\tau)$ often provide a useful alternative representation for an ellipse where $\zeta\,=\,re^{i(\tau\,+\,t_{+})}\,.$ Defining $\nu\,=\,\phi\,+\,2\tau\,,$ then
\begin{equation}\label{dotnu}
\dot{\nu}\;=\;\Delta\,+\,2\mid\sigma\mid\sin\nu\quad,\quad\frac{\dot{r}}{r}\;=\;\frac{1}{2}\theta\,-\,\mid\sigma\mid\cos\nu\quad.
\end{equation}
Setting $z\,=\,2(\tau\,+\,t_{+})\,,$ differentiating the first and using the second of eq.(\ref{dotnu}) then leads to the non-linear (inverted) tidal gravitational ``pendulum" equation $\ddot{z}\,+\,\frac{2\dot{r}}{r}\dot{z}\,-\,2\Psi\sin (z\,-\,\beta)\,=\,0$ \cite{Me2}. We also find that $r$ satisfies a harmonic oscillator-like equation $-\frac{\ddot{r}}{r}\,=\,\Phi_{00}\,+\,\Psi\cos (z\,-\,\mu)\,-\,\frac{1}{4}\dot{z}^{2}\,.$

\section{Gravitational Optics Equations}

With the new variables $\phi$ and $w\,,$ we obtain from eqs.(\ref{sq3}), (\ref{sq3a}) and (\ref{es1}) the following coupled system of non-linear ordinary differential equations, and one of the main results of this paper
\begin{eqnarray}
\dot{\phi} & = & \Delta\,+\, 2\mid\sigma\mid\sin\phi\coth w\quad,\label{sq5}\\
\dot{w} & = & -2\mid\sigma\mid\cos\phi\quad,\label{sq6}\\
2\dot{v} & = & \Delta\,+\,2\mid\sigma\mid\sin\phi\tanh\frac{1}{2}w\quad,\label{sq7}
\end{eqnarray}
where we introduced the ``detuning" $\Delta\,=\,\omega\,-\,\Omega$ and an additional angular variable $v\,=\,t_{-}\,+\,t_{+}\,-\,\frac{1}{2}\varphi\,=\,t_{0}\,+\,t\,+\,\varphi_{1}\,-\,\frac{1}{2}\varphi\,.$ Eqs.(\ref{sq5}) - (\ref{sq7}) resemble the two-mode squeezed state evolution equations obtained in the context of cosmological particle creation due to inflation \cite{Grishchuk}. Here, $\mid\sigma\mid$ is analogous to the $a'/a$ `pump' term in cosmology, $w$ is the squeeze parameter, $\phi$ the squeeze angle and $2v$ the rotation angle. In an inflationary context, the strong gravitational field in the early universe parametrically amplifies (equivalent to $w\,>\,0$) zero-point quantum oscillations of relic gravitational waves. Analogously, one might envisage an amplification of the gravitational field due to the tidal effects of the Weyl tensor. In the absence of tidal effects ($\Psi\,=\,0$) then $\dot{\phi}\,=\,\omega\,,$ $w\,=\,0$ and $2\dot{v}\,=\,\omega\,.$ Structurally similar equations to eqs.(\ref{sq5}) - (\ref{sq7}) were obtained in \cite{Frittelli}, \cite{Korotky} and \cite{Chakra} but these representations did not make a connection with other areas of physics, as far as the author is aware, which we will describe later in this paper.

To facilitate analytic and numerical solutions of eqs.(\ref{sq5}) - (\ref{sq7}), which are not discussed further in this paper, we can use the spatial distance travelled by a photon, $d\ell\,=\,\bar{\omega}d\lambda\,,$ where $\bar{\omega}\,=\,-k^{\mu}u_{\mu}$ is the frequency measured by an observer with 4-velocity $u^{\mu}\,,$ giving
\begin{eqnarray}
\phi' & = & \frac{\Delta}{\bar{\omega}}\,+\,\frac{2\mid\sigma\mid}{\bar{\omega}}\sin\phi\coth w\quad,\label{sq5c}\\
w' & = & -\frac{2\mid\sigma\mid}{\bar{\omega}}\cos\phi\quad,\label{sq6c}\\
2v'& = & \frac{\Delta}{\bar{\omega}}\,+\, \frac{2\mid\sigma\mid}{\bar{\omega}}\sin\phi\tanh\frac{1}{2}w\quad,\label{sq7c}
\end{eqnarray}
where $'$ denotes a derivative with respect to $\ell\,.$

Eqs (\ref{sq5}) and (\ref{sq6}) can be combined into a single complex differential equation upon introducing a complex angular variable $\Phi (\phi\,,\,w)$ which essentially combines the orientation of the elliptical image with its axis ratio
\begin{equation}\label{PHI}
\Phi (\phi\,,\,w)\;=\;\phi\,-\,iF(w)\quad,\quad \tanh\frac{1}{2}w\;=\;e^{-F(w)}\quad.
\end{equation}
Eqs (\ref{sq5}) and (\ref{sq6}) then become
\begin{equation}\label{Phiprime}
\Phi'\;=\;\frac{\Delta}{\bar{\omega}}\,+\,\frac{2\mid\sigma\mid}{\bar{\omega}}\sin\Phi\quad.
\end{equation}
At resonance $(\Delta\,=\,0)$ and $w\,\gg\,1\,,$ one solution to this equation is $\phi\,\simeq\,n\pi$ where $n$ is an integer. This is just a more interesting form of the ellipticity ($\epsilon$) evolution equation obtained in \cite{perlick}. In polarisation optics, eq.(29) is analogous to the Kravtsov–Orlov equation for the complex angle $\Theta\,=\,\Theta_{1}\,+\,i\Theta_{2}$, where $\Theta_{1}$ is a position angle and $\Theta_{2}$ determines the circular polarisation \cite{Kravtsov}.

\section{Hyperbolic Bloch Representation}

\subsection{Pseudospherical Coordinates}

Let us introduce complex amplitudes $\hat{\zeta}_{1,2}\,=\,\zeta_{1,2}\xi^{-1}\,,$ where $\dot{\xi}\xi^{-1}\,=\,-\rho$ such that $\xi\,=\,D_{a}e^{i\psi}$ and $\omega\,=\,2\dot{\psi}\,=\,2LD_{a}^{-2}\,.$ Eq.(\ref{zeta3b}) can then be written in the compact form 
\begin{equation}\label{sq1}
\dot{\hat{\zeta}}_{1,2}\;=\;-\sigma e^{-2i(t_{0}\,+\,\psi)}\hat{\zeta}^{*}_{2,1}\quad.
\end{equation}
Now define complex, time-dependent functions $\alpha\,=\,\hat{\zeta}_{1}e^{i\chi}$ and $\beta\,=\,\hat{\zeta}_{2}e^{i\chi} $ where $\chi\,=\,t_{0}\,+\,\psi\,-\,\frac{1}{2}\varphi\,.$ After a rescaling with $D_{a}\,=\,c_{0}\sqrt{C_{+}C_{-}}\,,$ $\alpha_{1}\,=\,\sqrt{2}c_{0}\alpha$ and $\beta_{1}\,=\,\sqrt{2}c_{0}\beta\,$, with $c_{0}$ independent of $\lambda\,,$ we find that
\begin{equation}\label{AB}
\alpha_{1}\;=\;e^{i\theta_{1}(\lambda)}\cosh\frac{1}{2}w\quad,\quad\beta_{1}\;=\;e^{i\theta_{2}(\lambda)}\sinh\frac{1}{2}w\quad,
\end{equation}
where $\theta_{1,2}$ are two real phases such that $\mid\alpha\mid^{2}\,-\,\mid\beta\mid^{2}\,=\,\frac{1}{2c_{0}^{2}}\,.$ The amplitudes $\alpha_{1}\,,\beta_{1}$ are somewhat like Bogoliubov coefficients. The evolution equations for $\alpha_{1}$ and $\beta_{1}$ have a Schr\"{o}dinger-like form with a complex ``state" vector (analogous to a qubit, two-level atom or polarised photon) $\mathbf{\Psi}\,=\,(\alpha_{1}\,,\,\beta_{1}^{*})^{T}$ and two-level Hamiltonian $\mathbf{H}$
\begin{equation}\label{Sch1}
i\dot{\mathbf{\Psi}}\;=\;\mathbf{H}\mathbf{\Psi}\quad,
\end{equation}
where
\begin{equation}\label{Ham1}
\mathbf{H}\;=\;\frac{1}{2}\begin{pmatrix}
-\Delta & -2i\mid\sigma\mid \\
-2i\mid\sigma\mid & \Delta
\end{pmatrix}\quad.
\end{equation}
This Hamiltonian is evidently non-Hermitian. We discuss the reason for this and its underlying group theoretical properties in the next sub-section.

Let us now define some coordinates $x_{i}\, (i\,=\,0\,,\,1\,,\,2)$
\begin{equation}
x_{0}\;=\;\mid\alpha_{1}\mid^{2}\,+\,\mid\beta_{1}\mid^{2}\quad,\quad  x_{1}\;=\;\alpha_{1}\beta_{1}\,+\,\beta^{*}_{1}\alpha_{1}^{*}\quad,\quad x_{2}\;=\;-i(\alpha_{1}\beta_{1}\,-\,\beta^{*}_{1}\alpha_{1}^{*})\quad.
\end{equation}
The coordinates $(x_{0}\,,\,x_{1}\,,\,x_{2})$ are those of a 3-vector $\mathbf{x}$ lying in the upper sheet (since $x_{0}\,>\,0$) $H_{2}^{+}$ of a two-sheet hyperboloid since
\begin{equation}\label{TSH}
x_{1}^{2}\,+\,x_{2}^{2}\,-\,x_{0}^{2}\;=\;-1\quad.
\end{equation}
The inner product between hyperbolic Bloch 3-vectors $\mathbf{a}$ and $\mathbf{b}$ on $H_{2}^{+}$ is defined as $\mathbf{a}.\mathbf{b}\,=\,-a_{0}b_{0}\,+\,a_{1}b_{1}\,+\,a_{2}b_{2}$ such that $\mathbf{x}.\mathbf{x}\,=\,-1\,.$ Using eq.(\ref{AB}), we obtain the representation
\begin{equation}\label{UV}
x_{0}\;=\;\cosh w\quad,\quad x_{1}\;=\;\sinh w\cos\phi\quad,\quad x_{2}\;=\;\sinh w\sin\phi\quad,
\end{equation}
where $w\,\geq\,0$ and $0\,\leq\,\phi\,\leq\,2\pi\,.$ Just like a spherical coordinate system, $w$ is the distance from an arbitrarily chosen origin and $\phi$ is the azimuth of a point measured from the polar axis such that the metric on $H_{2}^{+}$ has the form $ds^{2}\,=\,dw^{2}\,+\,\sinh^{2}w\, d\phi^{2}\,.$

Isometries of $H_{2}^{+}$ which preserve eq.(\ref{TSH}) are reflections across planes containing the $x_{0}$ axis and hyperbolic rotations about the $x_{2}$ axis
\begin{equation}
x_{0}\;=\;x'_{1}\sinh w'\,+\,x'_{0}\cosh w'\quad,\quad x_{1}\;=\;x'_{1}\cosh w'\,+\,x'_{0}\sinh w'\quad,\quad x_{2}\;=\;x'_{2}\label{HS}\quad.
\end{equation} 
The first equation is the hyperbolic law of cosines and the last is the hyperbolic law of sines such that $x_{0}'\,=\,\cosh w'\,,$ $x_{1}'\,=\,\sinh w'\cos\phi'$ and $x'_{2}\,=\,\sinh w'\sin\phi'.$ Physically, the hyperbolic law of cosines in this context is just a representation of the fact that an observed elliptical image may be the result of the effect of shear on light rays from an intrinsically elliptical source, such as a galaxy, which can itself be interpreted as resulting from the effect of a fictitious shear on a fictitious circular source \cite{KK}. In addition, the rotation about the $x_{0}$ axis isometry has the form
\begin{equation}
x_{0}\;=\;x'_{0}\quad,\quad x_{1}\;=\;x'_{1}\cos\chi_{1}\,-\,x'_{2}\sin\chi_{1}\quad,\quad x_{2}\;=\;x'_{1}\sin\chi_{1}\,+\,x'_{2}\cos\chi_{1}\quad.
\end{equation}
We will explore these isometries elsewhere in the context of Thomas precession in gravitational optics.

We now introduce two complex functions $f_{\pm}(\lambda)$ 
\begin{equation} \label{f1}
f_{\pm}\;=\;\alpha_{1}\,\pm\,\beta_{1}^{*}\;=\;e^{i\theta_{1}}\cosh \frac{1}{2}w\,\pm\,e^{-i\theta_{2}}\sinh\frac{1}{2}w\quad.
\end{equation}
Without too much difficulty, one can reproduce eqs.(\ref{sq5}) - (\ref{sq7}) if 
\begin{equation}\label{T1a}
\theta_{1}\,=\,t_{0}\,+\,\varphi_{1}\,-\,\frac{1}{2}\varphi\quad,\quad\theta_{2}\,=\,t_{0}\,+\,\varphi_{2}\,-\,\frac{1}{2}\varphi\quad 
\end{equation}
with $\phi\,=\,\theta_{1}\,+\,\theta_{2}$ up to $\pm 2\pi$ multiples. The complex functions $f_{\pm}$ evolve according to
\begin{equation}\label{fdot}
\dot{f}_{\pm}\;=\;\frac{1}{2}i\Delta f_{\mp}\,\mp\,\mid\sigma\mid f_{\pm}\quad.
\end{equation}
Defining $dx\,=\,\Delta d\lambda$ and $a\,=\,\frac{2\mid\sigma\mid}{\Delta}\,$, eqs.(\ref{fdot}) imply that $f_{\pm}$ satisfy complex harmonic oscillator equations
\begin{equation}\label{fho}
\frac{d^{2}f_{\pm}}{dx^{2}}\,+\,\omega_{\pm}^{2}(x)f_{\pm}\;=\;0\quad,
\end{equation}
where the frequencies $\omega_{\pm}(x)$ are determined by the Riccati equations $\omega_{\pm}^{2}\,=\,\frac{1}{4}\,\pm\,a'\,-\,a^{2}\,.$ 

In a sense, $f_{+}$ acts like a complex field coordinate and $f_{-}$ like a complex generalised conjugate momentum in a Schr\"{o}dinger-like picture. To see this, we define a complex generalised momentum $p_{+}\,=\,2\dot{f}_{+}\,+\,2\mid\sigma\mid f_{+}\,.$  By eq.(\ref{fdot}), then $p_{+}\,=\,i\Delta f_{-}\,.$ A corresponding Hamiltonian $H_{0}$ can be defined as $H_{0}\,=\,\frac{1}{2\Delta}(\mid p_{+}\mid^{2}\,+\,\Delta^{2}\mid f_{+}\mid^{2})\,=\,\Delta x_{0}$ on using eq.(\ref{f1}). Introducing another momentum variable $p_{+}'\,=\,2\dot{f}_{+}$ such that $p_{+}\,=\,p_{+}'\,+\,2\mid\sigma\mid f_{+}\,,$ then
\begin{equation}\label{H4}
H\;=\;\frac{1}{2\Delta}\left[\mid p'_{+}\mid^{2}\,+\,(\Delta^{2}\,-\,4\mid\sigma\mid^{2})\mid f_{+}\mid^{2}\right]\;=\;\Delta x_{0}\,+\,2\mid\sigma\mid x_{2}\quad.
\end{equation}
Thus, through the first term in eq.(\ref{H4}), $H$ has a standard harmonic oscillator-like form.

\subsection{Non-Hermitian Hamiltonian}

The non-Hermitian $(\mathbf{H}^{\dagger}\,\neq\, \mathbf{H})$ nature of $\mathbf{H}$ is not entirely unexpected given the underlying $SU(1,1)$ group structure associated with the hyperbolic geometry of our model. Unlike $SU(2)$ associated with the two-level atomic model, the $SU(1,1)$ group is non-compact, which means all its unitary irreducible representations are infinite-dimensional. We will consider below a non-unitary representation of $SU(1,1)$ with generators represented by non-Hermitian, finite-dimensional matrices.

The associated $SU(1,1)$ evolution matrix $\mathbf{U}$ also satisfies the Schr\"{o}dinger-like equation $i\dot{\mathbf{U}}\,=\,\mathbf{H}\mathbf{U}$ where
\begin{equation}
\mathbf{U}\;=\;\begin{pmatrix}
\alpha_{1} & \beta_{1} \\
\beta_{1}^{*} & \alpha_{1}^{*} \\
\end{pmatrix}\quad.
\end{equation}
Although non-unitary, $\mathbf{U}$ satisfies the $SU(1,1)$ condition $\mathbf{U}^{\ddagger}\mathbf{U}\,=\,\mathbf{I}\,,$ where $\mathbf{U}^{\ddagger}\,=\,\sigma_{z}\mathbf{U}^{\dagger}\sigma_{z}$ and $\mathbf{I}$ is the identity matrix. Since $\mid\alpha_{1}\mid^{2}\,-\,\mid\beta_{1}\mid^{2}\,=\,1$ and $\mid\alpha_{1}\mid^{2}\,\geq\,1\,,$ then $\mid\alpha_{1}\mid^{2}$ and $\mid\beta_{1}\mid^{2}$ cannot be interpreted as probabilities and $\mathbf{U}$ cannot be interpreted as a time evolution matrix. Notions of probability and statistics can be recovered, however, in the next sub-section when we introduce a particular form of density matrix.

Expressing $\mathbf{H}$ in terms of the generators $\{\boldsymbol{K}_{0}\,,\,\boldsymbol{K}_{+}\,,\,\boldsymbol{K}_{-}\}$ of the $su(1,1)$ Lie algebra gives
\begin{equation}\label{Hsu11}
\mathbf{H}\;=\;-\Delta\boldsymbol{K}_{0}\,+\,\nu^{*}(\lambda)\boldsymbol{K}_{+}\,+\,\nu (\lambda)\boldsymbol{K}_{-}\quad,
\end{equation}
where $\nu (\lambda)\,=\,-i\mid\sigma\mid\,=\,-\nu^{*}(\lambda)$ and
\begin{equation}
\boldsymbol{K}_{0}\;=\;\frac{1}{2}\begin{pmatrix}
1 & 0 \\
0 & -1 \\
\end{pmatrix}\quad,\quad \boldsymbol{K}_{+}\;=\;\begin{pmatrix}
0 & -1 \\
0 & 0 \\
\end{pmatrix}\quad,\quad \boldsymbol{K}_{-}\;=\;\begin{pmatrix}
0 & 0 \\
1 & 0 \\
\end{pmatrix}\quad,
\end{equation}
which satisfy the commutation relations $[\boldsymbol{K}_{0}\,,\,\boldsymbol{K}_{\pm}]\,=\,\pm\boldsymbol{K}_{\pm}$ and $[\boldsymbol{K}_{+}\,,\,\boldsymbol{K}_{-}]\,=\,-2\boldsymbol{K}_{0}\,.$ An alternative basis for the $su(1,1)$ algebra is $\boldsymbol{K}_{0}\,=\,\frac{1}{2}\boldsymbol{\sigma}_{z}\,,$ $\boldsymbol{K}_{1}\,=\,-\frac{i}{2}\boldsymbol{\sigma}_{y}$ and $\boldsymbol{K}_{2}\,=\,\frac{i}{2}\boldsymbol{\sigma}_{x}\,,$
where $\boldsymbol{\sigma}_{x,y,z}$ are the Pauli matrices and $\boldsymbol{K}_{\pm}\,=\,\boldsymbol{K}_{1}\,\pm\,i \boldsymbol{K}_{2}\,.$ In addition, the $su(1,1)$ Lie algebra can be realised in terms of bilinear products of the annihilation and creation operators of two boson modes. This agrees with our bosonic gravitational model.

Equation (\ref{Hsu11}) shows that $\mathbf{H}$ has the form $\mathbf{H}\,=\,\mathbf{H}_{0}\,-\,i\boldsymbol{\Gamma}$ where $\mathbf{H}_{0}\,=\,\mathbf{H}_{0}^{\dagger}\,=\,-\Delta\mathbf{K}_{0}$ and $\boldsymbol{\Gamma}\,=\,\boldsymbol{\Gamma}^{\dagger}\,=\,\mid\sigma\mid\boldsymbol{\sigma}_{x}\,.$ The matrix $\mathbf{H}_{0}$ is the Hermitian part generating unitary dynamics and $-i\boldsymbol{\Gamma}$ is the anti-Hermitian part governing gain and loss in the system. Furthermore, as an $SU(1,1)$ Hamiltonian, $\mathbf{H}$ is also pseudo-Hermitian, which means there exists a Hermitian matrix $\boldsymbol{\eta}$ such that $\boldsymbol{\eta}\mathbf{H}\boldsymbol{\eta}^{-1}\,=\,\mathbf{H}^{\dagger}$ \cite{Mosta}. Such $\boldsymbol{\eta}$ is not unique, although the simplest choice, consistent with the $SU(1,1)$ group structure, is $\boldsymbol{\eta}\,=\,\boldsymbol{\sigma}_{z}\,,$ where $\boldsymbol{\sigma}_{z}$ is a Pauli matrix. In a certain sense, pseudo-Hermitian Hamiltonians are a subclass of non-Hermitian Hamiltonians. The matrix $\mathbf{H}$ also has the form of a time-dependent Hamiltonian that meets all the conditions of $PT$ quantum mechanics where $PT\,=\,\boldsymbol{\sigma}_{z}$ \cite{JonesSmith}.

\subsection{Hyperbolic Bloch Equations}

In terms of the coordinates $x_{i}$ introduced in eq.(\ref{UV}), eqs (\ref{sq5}) - (\ref{sq7}) become another first-order coupled system of ordinary differential equations
\begin{eqnarray}
\dot{x}_{0} & = & -2\mid\sigma\mid x_{1}\quad, \label{Ndot} \\
\dot{x}_{1} & = & -2\mid\sigma\mid x_{0}\,-\,\Delta x_{2}\quad, \label{Udot} \\
\dot{x}_{2} & = & \Delta x_{1}\quad\label{Vdot}.
\end{eqnarray}
These equations are the hyperbolic Bloch equations of general relativity, the main result of this paper. Although the physics is different, they are analogous to the standard form of the optical Bloch equations for a two-level atomic system interacting with an electric field \cite{Eberly}. Hyperbolic Bloch equations also model the excited atoms of the Bose-Einstein condensate \cite{Kira} - see also \cite{Gibbon}. Indeed, eqs (\ref{Ndot}) - (\ref{Vdot}) can be expressed as a von Neumann-like equation $i\dot{\boldsymbol{\rho}}\,=\,[\boldsymbol{\rho}\,,\,\mathbf{H}]$ where we used eq.(\ref{Ham1}) and the density matrix
\begin{equation}
\boldsymbol{\rho}\;=\;\frac{1}{2}\begin{pmatrix}
1\,+\,ix_{0} & x_{2}\,+\,ix_{1} \\
x_{2}\,-\,ix_{1} & 1\,-\,ix_{0} \\
\end{pmatrix}\quad,
\end{equation}
such that $\mathrm{det}\boldsymbol{\rho}\,=\,\frac{1}{2}$ and $\mathrm{Tr}\,\boldsymbol{\rho}\,=\,1\,.$ Despite the non-Hermitian, albeit normalised, density matrix $\boldsymbol{\rho}$ of eq.(47), we introduce a (non-normalised) Hermitian, reduced density matrix $\boldsymbol{\varrho }\,=\,\boldsymbol{\Psi}^{T}\boldsymbol{\Psi}\,,$ which implies
\begin{equation}
\boldsymbol{\varrho}\;=\; \begin{pmatrix}
\mid\alpha_{1}\mid^{2} & \alpha_{1}\beta_{1} \\
\alpha_{1}^{*}\beta_{1}^{*} & \mid\beta_{1}\mid^{2} \\
\end{pmatrix}\;=\;\frac{1}{2}(\boldsymbol{\sigma}_{z}\,+\,\mathbf{x}.\boldsymbol{\sigma})\quad,
\end{equation}
where $\mathbf{x}\,=\,(x_{1}\,,\,x_{2}\,,\,x_{0})\,,$ $\boldsymbol{\sigma}\,=\,(\boldsymbol{\sigma}_{x}\,,\,\boldsymbol{\sigma_{y}}\,,\,\boldsymbol{\sigma}_{z})$ and the dot product is the standard Euclidean dot product. The matrix $\boldsymbol{\varrho}$ is related to $\boldsymbol{\rho}$ through $\boldsymbol{\varrho}\,=\,\frac{1}{2}(1\,+\,i)\mathbf{I}\,-\,i\sigma_{z}\boldsymbol{\rho}$ and $\mathrm{Tr}\,(\boldsymbol{\sigma}_{z}\boldsymbol{\varrho})\,=\,1\,.$ A probabilistic and statistical interpretation can be recovered from our non-Hermitian dynamics by introducing a normalised density matrix $\boldsymbol{\varrho}'\,=\,\frac{\boldsymbol{\varrho}}{\mathrm{Tr}\,\boldsymbol{\varrho}}$ whose trace is conserved. The matrix $\boldsymbol{\varrho}'$ evolves according to a Lindblad-like equation familiar in the theory of open quantum systems $i\dot{\boldsymbol{\varrho}}'\,=\,[\mathbf{H}_{0}\,,\,\varrho]\,-\,i\{\boldsymbol{\Gamma}\,,\,\varrho\}\,+\,2i\boldsymbol{\varrho}'\mathrm{Tr}\,(\boldsymbol{\Gamma}\boldsymbol{\varrho}')\,,$ where $\{\,\}$ denotes the anti-commutator and $\mathbf{H}_{0}$ and $\boldsymbol{\Gamma}$ were introduced in the previous section - see \cite{Sergi}. 

Eq.(\ref{TSH}) can be expressed in a form analogous to an interacting Bose gas \cite{Kira} \cite{Gibbon}
\begin{equation}
(1\,+\,2N_{+})^{2}\,-\,\mid P\mid^{2}\;=\;1\quad,
\end{equation}
where $N_{+}$ denotes an ``occupation number" such that $x_{0}\,=\,1\,+\,2N_{+}$ and $P\,=\,x_{2}\,+\,ix_{1}$ denotes a ``polarisation" function. The appearance of the $1\,+\,2N_{+}$ term here perhaps reflects the underlying bosonic nature of the gravitational field. The polarisation and occupation number evolve according to 
\begin{equation}
i\dot{P}\;=\;\Delta P\,+\,2\mid\sigma\mid (1\,+\,2N_{+})\quad,\quad\dot{N}_{+}\;=\; 2\mid\sigma\mid\mathrm{Im}\,P^{*}\quad.
\end{equation}
These are the hyperbolic analogues of the semiconductor  Bloch equations \cite{Lindberg}.

Eqs (\ref{Ndot}) - (\ref{Vdot}) can also be expressed in matrix form as
\begin{equation}
\begin{pmatrix}
\dot{x}_{0} \\
\dot{x}_{1} \\
\dot{x}_{2} 
\end{pmatrix}\;=\;\begin{pmatrix}
0 & -2\mid\sigma\mid & 0 \\
-2\mid\sigma\mid & 0 & -\Delta \\
0 & \Delta & 0
\end{pmatrix} \begin{pmatrix}
x_{0} \\
x_{1} \\
x_{2}
\end{pmatrix}\quad,
\end{equation}
or in vector form like a classical spinning top or gyroscope driven by a rotation vector $\mathbf{\Omega}$
\begin{equation}\label{dots}
\dot{\mathbf{x}}\;=\;\mathbf{\Omega}\,\tilde{\times}\,\mathbf{x}\quad,
\end{equation}
where 
\begin{equation}\label{R}
\mathbf{\Omega}\;=\;(\Delta\,,\,0\,,\,-2\mid\sigma\mid)^{T}\quad,
\end{equation}
and $\tilde{\times}$ denotes a vector product in $H_{2}^{+}$ (such that if $\mathbf{X}\,=\,(x_{0}\,,\,x_{1}\,,\,x_{2})$ and $\mathbf{Y}\,=\,(y_{0}\,,\,y_{1}\,,\,y_{2})$ are two vectors in $H_{2}^{+}$ then $\mathbf{X}\,\tilde{\times}\,\mathbf{Y}\,=\,(-x_{1}y_{2}\,+\,x_{2}y_{1}\,,\, x_{2}y_{0}\,-\,x_{0}y_{2}\,,\, x_{0}y_{1}\,-\,x_{1}y_{0})$). This is another key result of this paper. Eq.(\ref{dots}) describes the precessional motion of $\mathbf{x}\,,$ analogous to the electric pseudospin or Stokes vector, about $\mathbf{\Omega}(\lambda)$ on $H_{2}^{+}$ with angular velocity
\begin{equation}\label{midomega}
\mid\mid\mathbf{\Omega}\mid\mid\;=\;\pm\sqrt{\Delta^{2}\,-\,(2\mid\sigma\mid)^{2}}\quad.
\end{equation}
In the spirit of \cite{Aravind}, it might be more accurate to describe the hyperbolic precession as a pseudo-precession since the $SU(1,1)$ orbits are open whilst those for $SU(2)$ are closed. In general relativity, therefore, the end points of the hyperbolic vector $\mathbf{x}$ move about the vector $\mathbf{\Omega}$ under a torque $\mathbf{\Omega}\,\tilde{\times}\,\mathbf{x}\,$. The magnitude $\mid\mid\mathbf{\Omega}\mid\mid$ is a hyperbolic generalised (on account of the detuning $\Delta$) Rabi frequency in a two-level atom interpretation. An $SU(1,1)$ precession model was also explored in \cite{Dattoli}.

In polarisation optics, Stokes parameters form a vector whose motion is described on the surface of the Poincar\'{e} sphere. Its poles represent circulary polarised light, the equator represents linearly polarised light and positions around the sphere represents the angle of polarisation. A birefringence vector, which represents the anisotropy of the medium, precesses about this Stokes vector \cite{Kubo}\cite{Kubo1}. One might, therefore, interpret $\mathbf{\Omega}$ as a hyperbolic birefringence 3-vector on $H_{2}^{+}\,$.

We note that $\mid\mid\mathbf{\Omega}\mid\mid $ represents the eigenvalues of the Hamiltonian matrix $\mathbf{H}$ in eq.(\ref{Ham1}). The eigenvalues of $\mathbf{H}$ are real if $\frac{1}{4}\Delta^{2}\,\geq\,\mid\sigma\mid^{2}\,.$ For a $PT$-symmetric Hamiltonian, either all its eigenvalues are real (unbroken $PT$-symmetry) or some are real and some are complex (broken $PT$-symmetry) \cite{Bender1}. To determine whether a phase transition from an unbroken to a broken $PT$ phase can occur, the key factor is the magnitude of the Weyl scalar $\Psi_{0}\,,$ the source of shear.

Whilst Bloch vectors were originally conceived in the context of nuclear magnetic resonance and the behaviour of a nuclear spin in a magnetic field under the influence of radio frequency pulses \cite{bloch}, they also represent semi-classical states of two-level atoms or qubits in quantum computing which have a geometric representation on the Bloch sphere. The interaction of an oscillatory classical electric field $\mathbf{E}(\mathbf{r}\,,\,t)$ with an atom - a light-matter interaction - can be described via the time dependent, semi-classical dipole interaction Hamiltonian $H_{I}\,=\,-\mathbf{d}.\mathbf{E}\,,$ where $\mathbf{d}$ denotes the electric dipole moment and $\mathbf{r}$ is the position of the atom \cite{Eberly}. The dipole moment is defined by $\mathbf{d}\,=\,-e\hat{\mathbf{r}}$ where $-e$ denotes the electron charge and $\hat{\mathbf{r}}$ is the position vector of the electron relative to the nucleus so that $H_{I}\,=\,e\mathbf{\hat{r}}.\mathbf{E}$. The total Hamiltonian can be written $\mathcal{H} (\mathbf{r}\,,\,t)\,=\,H_{0}\,+\,H_{I} (\mathbf{r}\,,\,t)\,,$ where $H_{0}$ is the Hamiltonian for the unperturbed atom. Similarly, eq.(\ref{H4}) can be also written in the compact form $H\,=\,-\mathbf{\Omega}.\mathbf{x}\,\neq\,,$ constant with the dot product on $H^{+}_{2}\,,$. Alternatively, we can express the non-Hermitian Hamiltonian eq.(33) as $\mathbf{H}\,=\,-\,\frac{1}{2}\Delta\,\boldsymbol{\sigma}_{z}\,-\,i\mid\sigma\mid\boldsymbol{\sigma}_{x}\,=\,\boldsymbol{\Omega}.\mathbf{K}$ (with the $H_{2}^{+}$ dot product) where $\boldsymbol{\Omega}$ is defined in eq.(52) and $\mathbf{K}\,=\,(\mathbf{K}_{0}\,,\,\mathbf{K}_{1}\,,\,\mathbf{K}_{2})\,.$ This is analogous to the interaction of the spin angular momentum with a magnetic field, which has the familiar form $H_{S}\,=\,-\gamma\mathbf{B}.\mathbf{S}$ (Euclidean dot product), where $\mathbf{B}$ is the magnetic field vector, $\gamma$ the gyromagnetic ratio and $\mathbf{S}\,=\,\frac{1}{2}\hbar\boldsymbol{\sigma}$ the spin angular momentum.

\section{Conclusion}

This paper has established a connection between general relativity and two-level atomic physics through hyperbolic geometry. We introduced a phase $\phi$ and axis ratio parameter $w\,,$ related to the properties of elliptical cross sections of a congruence of light rays, which were interpreted as polar coordinates on the upper sheet of a two-sheet hyperboloid $H_{2}^{+}\,$. Their evolution equations were driven by the shear and twist in the light rays. We transformed the scalar evolution equations into a single vector equation describing the precession of a hyperbolic Bloch about a rotation or birefringence vector. The magnitude of the complex shear rate played a similar role to the electric field in the two-level atomic theory. An interpretation in terms of polarisation states on $H^{+}_{2}$ may be possible in the context of \cite{Giust} where it was shown that any state of light propagating inside a multi-layer can be represented as a point on $H_{2}^{+}\,$. It is tempting to speculate that the connection between the shear rate appearing in the hyperbolic Bloch vectors and the electric field appearing in the optical Bloch equations is related to the fact that in higher-order geometrical optics in curved spacetime, the first-order corrections to the electric field are determined by the shear \cite{Anile}, \cite{Ehlers},\cite{Dolan}, \cite{DK}.

Our gravitational optics model is a two-level system described by a time-dependent non-Hermitian $SU(1,1)$ Hamiltonian, similar to the model investigated in \cite{Grimaudo}, \cite{Grimaudo1}, which arose from a Schr\"{o}dinger-like representation of the hyperbolic Bloch equations. Whilst the dynamics is non-unitary, indicative of an open system, we constructed in the $SU(1,1)$ theory a density matrix whose trace is conserved. Given the great interest in non-Hermitian Hamiltonians following the work of Bender and Boettcher in the context of $PT$-quantum mechanics \cite{Bender}, it is enlightening to have established in this paper a general relativistic connection with this theory.

For discussion elsewhere is the analogy between the formalism of Schwinger particle production due to strong electric fields and the behaviour of light rays in curved spacetime. To see this, the formal solutions to eqs.(\ref{Udot}) and (\ref{Vdot}) for $x_{1}$ and $x_{2}$ are
\begin{eqnarray}
x_{1}(\lambda) & = & -2\int^{\lambda}_{\lambda_{0}}d\lambda'\mid\sigma (\lambda')\mid x_{0}(\lambda')\cos\gamma (\lambda\,,\,\lambda')\quad,\\
x_{2}(\lambda) & = & 2\int^{\lambda}_{\lambda_{0}}d\lambda'\mid\sigma (\lambda')\mid x_{0}(\lambda')\sin\gamma (\lambda\,,\,\lambda')\quad,
\end{eqnarray}
where $\gamma (\lambda\,,\,\lambda')\,=\,\varphi(\lambda)\,-\,\varphi (\lambda')\,-\,\psi (\lambda)\,+\,\psi (\lambda')$ such that $\dot{\gamma}\,=\,-\Delta\,.$ Eq.(\ref{Ndot}) can, therefore, be expressed formally as an integro-differential kinetic- or Vlasov-like equation similar to the one obtained for bosonic pair production in flat spacetime due to an oscillating electric field \cite{schmidt}
\begin{equation}\label{dotN1a}
\dot{N}_{+}\;=\;2\mid\sigma\mid\int^{\lambda}_{\lambda_{0}}d\lambda'\mid\sigma (\lambda')\mid\left(1\,+\,2N_{+}(\lambda')\right)\cos\gamma (\lambda\,,\,\lambda')\quad,
\end{equation}
where $x_{0}\,=\,1\,+\,2N_{+}\,.$ What is interesting about this equation is the non-Markovian character of the source term since: (a) the change in $N_{+}$ at the point $\lambda$ depends on its entire ``history" of the light rays from the point $\lambda_{0}$ up to $\lambda\,,$ indicative of memory effects; (b) the integrand is a non-local function due to the $\cos\gamma (\lambda\,,\,\lambda') $ term which induces high-frequency oscillations. Approximate solutions of eq.(\ref{dotN1a}) in the non-Markovian and low-density limits $(N_{+}\,\ll\,1)$ will be obtained elsewhere.

Additional results on the connections between gravitational optics and hyperbolic geometry will be presented in future papers. In particular, we will consider Thomas precession in gravitational optics and the behaviour of trajectories on $H_{2}^{+}$ and their mapping to the hyperbolic plane and Poincar\'{e} disk. Also considered elsewhere is the theory of photoelasticity in curved spacetime and the associated Kuske and Neumann equations, which eqs (\ref{sq5}) - (\ref{sq7}) resemble. Exact solutions of the second-order ordinary differential equation for $\alpha_{1}\,,$ $\beta_{1}$ in eq.(\ref{AB}) via a mapping to the hypergeometric equation will be explored elsewhere.

\section{Acknowledgements}

The author is very grateful to Professor Valerio Faraoni for his assistance with the publication of this paper.

\end{document}